\documentclass[conference]{IEEEtran}

\usepackage{amsmath,amssymb}
\usepackage{graphicx}
\usepackage{booktabs}
\usepackage{array}
\usepackage{cite}
\usepackage{url}
\usepackage{microtype}
\usepackage{balance}

\graphicspath{{figures/}}
\newcommand{\bx}{\mathbf{x}}
\newcommand{\bs}{\mathbf{s}}
\newcommand{\bw}{\mathbf{w}}
\newcommand{\br}{\mathbf{r}}
\newcommand{\bP}{\mathbf{P}}
\newcommand{\bu}{\mathbf{u}}
\newcommand{\bW}{\mathbf{W}}
\newcommand{\be}{\mathbf{e}}
\newcommand{\bk}{\mathbf{k}}
\newcommand{\bI}{\mathbf{I}}
\newcommand{\bR}{\mathbf{R}}

\begin{document}

\title{Reference-Based Recursive Least-Squares Mitigation of Real Interference in Stereo Audio Recordings}

\author{\IEEEauthorblockN{Necati Kagan Erkek and Y. Ugur Ozcan}
\IEEEauthorblockA{Telecommunications Engineering, Department of Electronics, Information and Bioengineering\\
Politecnico di Milano, Piazza Leonardo da Vinci 32, 20133 Milan, Italy\\
necatikagan.erkek@mail.polimi.it, yilmazugur.ozcan@mail.polimi.it}}

\maketitle

\begin{abstract}
Reference-based adaptive interference cancellation is evaluated for stereo audio recordings corrupted by real train noise and environmental background. The observed signal is modeled as a clean stereo program contaminated by an additive disturbance generated by an external acoustic source through unknown propagation paths. A second stereo recording, representing another filtered observation of the same physical noise source, is used as the reference input of a multi-reference recursive least-squares (RLS) estimator. The estimated train-interference component is subtracted from the noisy audio and followed by a finite-impulse-response low-pass postfilter. Three 74.01 s real audio sequences sampled at 11.025 kHz are processed under identical algorithmic parameters. Since clean ground truth is not available, performance is assessed with no-reference indicators: waveform behavior, Welch spectral estimates, RMS change, and residual normalized correlation with the reference. With 30 taps per reference channel, 15 anti-causal taps, and forgetting factor 0.999, the maximum reference correlation is reduced from 0.386--0.832 before processing to 0.011--0.016 after processing. The corresponding correlation-ratio reduction is approximately 30.6--34.1 dB, while the output RMS decreases by 1.8--4.8 dB depending on section and stereo channel. The results demonstrate that real train interference, including environmental acoustic effects, can be substantially attenuated when a correlated reference recording is available.
\end{abstract}


\section{Introduction}
Interference mitigation remains a fundamental task in statistical signal processing, acoustic monitoring, communications, and audio enhancement \cite{spagnolini}. Audio recordings provide an informative experimental domain because algorithmic behavior can be examined through numerical metrics, waveform evolution, time-frequency representations, and direct listening. The workshop task defines a stereo program signal affected by an external noise source and provides a second stereo measurement related to the same source. The objective is to estimate a noise-mitigated stereo signal suitable for perceptual inspection using standard audio playback \cite{workshop}.

The acoustic material is based on real sound rather than a synthetically generated disturbance. The interference therefore includes actual train pass-by sound, environmental background, microphone coloration, propagation delay, possible reverberation, and time-varying spectral structure. Railway noise can contain low-frequency rolling components, broadband wheel-rail friction, wheel–track impacts, braking or traction events, mechanical transients, and local environmental reflections. These components vary with train speed, source-to-microphone distance, track or station geometry, and surrounding surfaces. Such effects generate amplitude modulation, intermittent bursts, and spectral overlap with speech or music-like content, which are difficult to reproduce with stationary laboratory noise. Consequently, synthetic white or colored noise often fails to represent the nonstationary character of transportation environments. A protocol based on real recordings is therefore valuable for future studies on robust acoustic sensing, railway-environment monitoring, assistive listening, transport communication systems, and practical audio restoration. 

Reference-based adaptive noise cancellation is appropriate when an auxiliary measurement is correlated with the unwanted component but ideally uncorrelated with the desired program \cite{widrow,kuo}. The reference does not need to be identical to the disturbance inside the target signal. Instead, an adaptive filter estimates the transfer path from the reference measurement to the interference observed in the program. The residual error after subtraction becomes the desired-signal estimate. The principle has been used in acoustic noise control, echo cancellation, and signal enhancement. \cite{haykin,sayed,diniz,elliott}.

The recursive least-squares algorithm is selected because of its fast convergence and strong tracking capability under correlated inputs. Compared with least-mean-square algorithms, RLS minimizes an exponentially weighted least-squares criterion and often reaches a useful solution with fewer samples, at the cost of higher arithmetic complexity and sensitivity to numerical conditioning \cite{sayed,diniz,cioffi}. In offline audio restoration, the additional complexity is acceptable because the signal can be processed after recording and the filter can exploit a carefully chosen tap structure.

The analysis provides a reproducible conference-style evaluation of reference-based mitigation of real train interference in stereo audio. The processing chain combines integer-delay alignment based on cross-correlation, a two-reference/two-output RLS noise estimator, and a conservative low-pass postfilter. No-reference interpretation is emphasized: because the clean stereo program is not supplied, true SNR and mean-square error cannot be claimed. Instead, residual reference correlation, RMS behavior, waveform changes, and spectral evidence characterize interference suppression under real environmental acoustic conditions. The real-sound protocol supplies a practical baseline for later studies involving controlled mixtures, subjective tests, and causal real-time implementations.

\section{Signal Model}
The mathematical model establishes the link between the physical railway-noise source, the primary stereo recording, and the auxiliary stereo reference. Both microphone paths are interpreted as unknown filters excited by the same real train source, while only the primary channel also contains the desired stereo program. This distinction is crucial for reference-based cancellation: the reference is informative because it is correlated with the disturbance, but direct subtraction is invalid because the paths differ in delay, magnitude response, spatial mixing, and environmental reflections.

\begin{figure}[h!]
    \centering
    \includegraphics[width=0.98\linewidth]{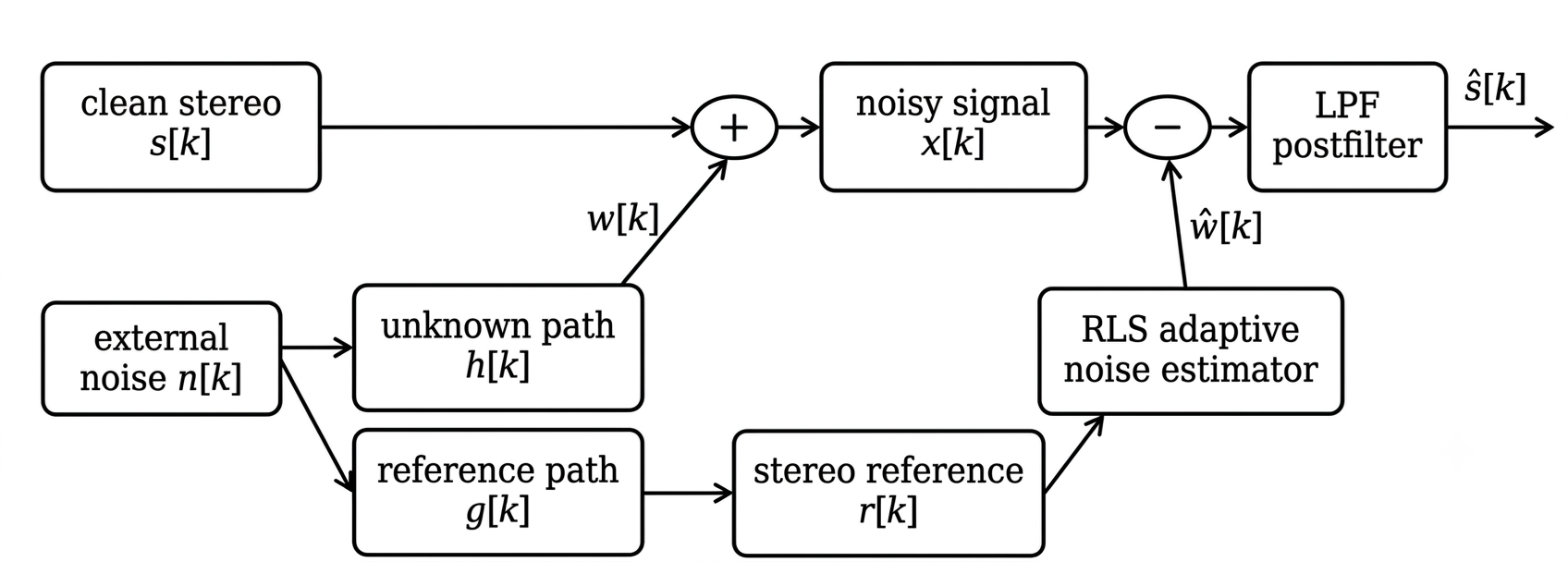}
    \caption{Reference-based adaptive mitigation model for real train interference in stereo audio. The clean stereo signal \(\bs[k]\) is corrupted by the interference \(\bw[k]\), where the external train-noise source \(n[k]\) propagates through an unknown acoustic path \(\mathbf{h}[k]\). A separate stereo reference \(\br[k]\) is generated by another path \(\mathbf{g}[k]\) from the same source. The RLS estimator predicts the reference-correlated interference \(\widehat{\bw}[k]\), the subtraction node produces the residual audio, and the low-pass postfilter generates the final estimate \(\widehat{\bs}[k]\).}
    \label{fig:model}
\end{figure}

Let the clean stereo audio at discrete time \(k\) be
\begin{equation}
    \bs[k] = \begin{bmatrix}s_1[k] & s_2[k]\end{bmatrix}^{T},
\end{equation}
and let the observed noisy stereo signal be
\begin{equation}
    \bx[k] = \bs[k] + \bw[k],
    \label{eq:model}
\end{equation}
where \(\bw[k]\) denotes the additive train-interference component. According to the workshop specification, the disturbance is generated by an external noise source \(n[k]\) through an unknown stereo path \(\mathbf{h}[k]\), while the reference recording is generated through another unknown stereo path \(\mathbf{g}[k]\) \cite{workshop}:
\begin{equation}
    \bw[k] = \mathbf{h}[k] * n[k], \qquad
    \br[k] = \mathbf{g}[k] * n[k].
    \label{eq:filteredrefs}
\end{equation}
The convolution operator \(*\) represents acoustic propagation and microphone-path filtering. Both \(\bw[k]\) and \(\br[k]\) are therefore related to the same source but are not equal in amplitude, delay, or spectral shape.

Fig.~\ref{fig:model} clarifies the role of each block in the cancellation chain. The unknown path \(\mathbf{h}[k]\) transforms the external train source before it appears in the noisy stereo signal, whereas \(\mathbf{g}[k]\) transforms the same source before reference acquisition. The RLS estimator therefore learns a path-compensated prediction of the reference-correlated disturbance rather than subtracting the reference directly. The postfilter is positioned after subtraction because residual high-frequency artifacts may remain when the real acoustic path is only approximately represented by a finite-length linear model.

The desired estimator has the form
\begin{equation}
    \widehat{\bs}[k] = \mathcal{P}\left\{\bx[k] - \widehat{\bw}[k]\right\},
    \label{eq:estimator}
\end{equation}
where \(\widehat{\bw}[k]\) is the adaptive estimate of the train-noise contribution and \(\mathcal{P}\{\cdot\}\) denotes the optional postfilter. Equation (\ref{eq:estimator}) highlights an important limitation: only the interference component correlated with the reference can be estimated. Noise that is uncorrelated with the reference, or desired audio that is correlated with the reference, cannot be perfectly separated by a linear reference-based method.

\section{Theoretical Background}
\subsection{Adaptive Noise Cancellation Principle}
Adaptive noise cancellation assumes that the primary input contains a desired signal plus interference, whereas the reference input contains information about the interference. The desired component should be statistically independent, or at least weakly correlated, with the reference. Under the condition, the reference can be filtered to synthesize the interference contribution without intentionally modeling the desired audio. Let \(z[k]\) be the scalar primary signal and let \(\bu[k]\) be a vector of reference samples. The standard adaptive cancellation model is
\begin{equation}
    e[k] = z[k] - \mathbf{c}^{T}[k-1]\bu[k],
    \label{eq:scalaranc}
\end{equation}
where \(e[k]\) is the residual and \(\mathbf{c}[k]\) is the adaptive coefficient vector. If the reference is uncorrelated with the desired signal, minimizing the output power tends to minimize the interference power in the residual \cite{widrow,haykin}. The same principle extends naturally to stereo output by estimating both program channels from the same reference vector.

The stereo case requires a matrix-valued adaptive filter. For two reference channels and two target channels, each target channel can depend on both reference channels because the unknown acoustic path is generally cross-coupled. Consequently, a two-input/two-output formulation is more appropriate than two independent single-reference cancellers. The multi-output residual is
\begin{equation}
    \be[k] = \bx[k] - \bW^{T}[k-1]\bu[k],
    \label{eq:mimoerr}
\end{equation}
where \(\bW[k]\in\mathbb{R}^{M\times 2}\) contains two adaptive coefficient vectors and \(M\) is the regressor dimension. The residual \(\be[k]\) is the RLS estimate of the stereo clean signal before postfiltering. The matrix formulation is important for stereo recordings because a train component captured by the left reference channel can still contribute to the right program channel after acoustic propagation, and vice versa.

\subsection{Exponentially Weighted Least-Squares Criterion}
RLS is obtained by minimizing an exponentially weighted least-squares cost. For the matrix-valued stereo filter, the objective at time \(k\) is
\begin{equation}
    J_k(\bW)=\sum_{i=1}^{k}\lambda^{k-i}
    \left\|\bx[i]-\bW^{T}\bu[i]\right\|_2^2
    + \delta \left\|\bW\right\|_F^2,
    \label{eq:cost}
\end{equation}
where \(0<\lambda\leq 1\) is the forgetting factor, \(\delta>0\) regularizes the initial inverse correlation matrix, and \(\|\cdot\|_F\) is the Frobenius norm. The parameter \(\lambda\) controls the memory of the estimator: values close to one exploit long data records, whereas lower values allow faster adaptation to time-varying conditions \cite{sayed,diniz}.

The normal-equation solution can be written as
\begin{equation}
    \bW[k] = \bR^{-1}[k]\mathbf{D}[k],
\end{equation}
where
\begin{align}
    \bR[k] &= \sum_{i=1}^{k}\lambda^{k-i}\bu[i]\bu^{T}[i] + \delta\lambda^k\bI, \\
    \mathbf{D}[k] &= \sum_{i=1}^{k}\lambda^{k-i}\bu[i]\bx^{T}[i].
\end{align}
Direct inversion of \(\bR[k]\) at every sample is computationally inefficient. The RLS recursion updates the inverse correlation matrix \(\bP[k]=\bR^{-1}[k]\) recursively through the matrix inversion lemma \cite{haykin,cioffi}:
\begin{align}
    \bk[k] &= \frac{\bP[k-1]\bu[k]}{\lambda + \bu^{T}[k]\bP[k-1]\bu[k]}, \label{eq:gain}\\
    \be[k] &= \bx[k] - \bW^{T}[k-1]\bu[k], \label{eq:err}\\
    \bW[k] &= \bW[k-1] + \bk[k]\be^{T}[k], \label{eq:wupdate}\\
    \bP[k] &= \lambda^{-1}\left(\bP[k-1] - \bk[k]\bu^{T}[k]\bP[k-1]\right). \label{eq:pupdate}
\end{align}
The initialization is \(\bW[0]=\mathbf{0}\) and \(\bP[0]=\bI/\delta\). Equations (\ref{eq:gain})--(\ref{eq:pupdate}) provide the complete adaptive estimator used for both stereo channels.

\subsection{Delay Alignment and Offline Tap Structure}
Physical acquisition paths introduce different propagation delays between the train source, the target microphones, and the reference microphones. A reference delay is estimated using the normalized cross-correlation
\begin{equation}
    \rho_{xr}[\ell] =
    \frac{\sum_k x[k]r[k-\ell]}
    {\sqrt{\sum_k x^2[k]}\sqrt{\sum_k r^2[k-\ell]}}.
    \label{eq:corr}
\end{equation}
The delay used in the experiments is the lag that maximizes the average absolute correlation across stereo channels within \(|\ell|\leq 200\) samples. The resulting corrections are small, but delay compensation improves the ability of a finite-length adaptive filter to model the reference-to-interference path.

The filter uses 30 taps per reference channel. Because the data are processed offline, the regressor includes both delayed and advanced reference samples. With 15 anti-causal taps, the tap offsets are
\begin{equation}
    \mathcal{T}=\{-14,-13,\ldots,0,\ldots,15\}.
\end{equation}
For reference channel \(q\in\{1,2\}\),
\begin{equation}
    \br_q[k]=\begin{bmatrix}r_q[k-14] & \cdots & r_q[k+15]\end{bmatrix}^{T}.
\end{equation}
The full regressor is obtained by stacking both reference-channel tap vectors:
\begin{equation}
    \bu[k]=\begin{bmatrix}\br_1^{T}[k] & \br_2^{T}[k]\end{bmatrix}^{T}.
\end{equation}
The regressor dimension is therefore \(M=60\). An anti-causal structure is not directly implementable without latency in real-time systems; however, it is suitable for offline restoration. A causal real-time version can be obtained by delaying the primary signal or by removing advanced reference samples.

\subsection{Low-Pass Postfilter}
After adaptive cancellation, residual artifacts can remain because the reference path may not perfectly span the disturbance path. A symmetric finite-impulse-response low-pass filter is therefore applied to the RLS residual. For order \(N_f=160\), the impulse response is a Hamming-windowed sinc design \cite{oppenheim,proakis}:
\begin{equation}
    p[n] = \frac{\sin\left(2\pi f_c(n-N_f/2)/f_s\right)}
    {\pi (n-N_f/2)}w_H[n],
\end{equation}
with appropriate limiting value at \(n=N_f/2\), normalization to unit DC gain, and cutoff \(f_c=0.55(f_s/2)\). For \(f_s=11025\) Hz, the cutoff is approximately 3.03 kHz. The postfilter suppresses upper-band residuals but can also attenuate desired high-frequency audio content. Therefore, the postfilter is interpreted as a conservative smoothing stage rather than an optimal perceptual enhancement method.

\section{Experimental Protocol}
\subsection{Dataset and Real Acoustic Conditions}
The dataset consists of three MATLAB files denoted A, B, and C. Each file contains a noisy stereo signal, a stereo reference measurement, and the sampling frequency. The sections contain 816000 samples per channel, corresponding to 74.01 s at 11.025 kHz. The same processing parameters are used for all sections to ensure comparable behavior across noise conditions.

The use of real sound is central to the experimental value. The recordings include a real train-noise component and environmental background rather than an artificial additive noise sequence. As a result, the algorithm must handle nonstationary acoustic structure, propagation-path mismatch, background ambience, and possible spectral overlap between the desired audio and the train disturbance. The characteristics make the task more representative of future applied studies than a purely synthetic benchmark.

\begin{table}[!t]
\centering
\caption{Processing parameters used for the three real stereo sections.}
\label{tab:params}
\begin{tabular}{ll}
\toprule
Parameter & Value \\
\midrule
Sampling frequency & 11025 Hz \\
Samples per channel & 816000 \\
Duration & 74.01 s \\
RLS forgetting factor, \(\lambda\) & 0.999 \\
RLS initialization, \(\delta\) & 0.1 \\
Reference channels & 2 \\
Taps per reference channel & 30 \\
Total coefficients per output & 60 \\
Anti-causal taps per reference & 15 \\
Lag search interval & \(\pm 200\) samples \\
Low-pass FIR order & 160 \\
Low-pass cutoff & 0.55 of Nyquist \\
\bottomrule
\end{tabular}
\end{table}

\subsection{Implementation Steps}
The full processing sequence is as follows. First, the channel-wise mean is removed from the noisy input and reference to avoid adapting to DC offsets. Second, cross-correlation is computed over the initial analysis segment and the reference is shifted by the estimated integer lag. Third, the two-channel reference regressor is formed at each sample and the RLS recursion estimates the reference-correlated interference. Fourth, the estimated interference is subtracted from both noisy channels. Fifth, the residual stereo signal is passed through the FIR low-pass postfilter and peak-normalized only for writing audio files. Quantitative metrics use the non-normalized processed signals.

The parameter choices follow standard adaptive-filtering practice. A large forgetting factor is selected because the recordings are relatively long and the objective is stable interference reduction rather than rapid tracking of a moving source. The tap length provides enough degrees of freedom for short acoustic path mismatch without making the inverse correlation matrix excessively large. The anti-causal part accounts for small alignment errors and symmetric offline modeling around the current sample. Table~\ref{tab:params} summarizes the configuration.

\subsection{Evaluation Metrics Without Clean Ground Truth}
Classical objective metrics such as SNR improvement, mean-square error, and source-to-distortion ratio require access to the clean signal \(\bs[k]\). The dataset does not provide clean ground truth, so such metrics would be unsupported. Evaluation is therefore based on no-reference quantities and visual evidence.

The first metric is the RMS change between the noisy input and the processed output:
\begin{equation}
    \Delta_{\mathrm{RMS},m}=20\log_{10}
    \frac{\mathrm{rms}(\widehat{s}_m)}{\mathrm{rms}(x_m)},
    \label{eq:rmsmetric}
\end{equation}
where \(m\in\{1,2\}\) denotes the stereo channel. RMS reduction is not equivalent to quality improvement because desired audio may also contribute to the RMS. Nevertheless, in the presence of strong additive train interference, RMS decrease provides a useful indication of attenuation strength.

The second metric is the maximum absolute normalized correlation between an output signal \(z_m[k]\) and the corresponding reference channel over the same lag window:
\begin{equation}
    \eta_m(z)=\max_{|\ell|\leq 200}\left|\rho_{z_m r_m}[\ell]\right|.
    \label{eq:etametric}
\end{equation}
The correlation-ratio reduction is
\begin{equation}
    C_m=20\log_{10}\frac{\eta_m(\widehat{s})}{\eta_m(x)}.
    \label{eq:corrsupp}
\end{equation}
Lower \(\eta_m(\widehat{s})\) indicates that less reference-related train noise remains in the processed signal. Spectral behavior is evaluated with Welch power spectral density estimates \cite{welch}. Listening-based assessment would be a valuable extension, preferably using standardized subjective procedures when sufficient listeners are available \cite{loizou,itur}.

\section{Results}
\subsection{Input Structure and Reference Relationship}
\begin{figure}[!t]
    \centering
    \includegraphics[width=\columnwidth]{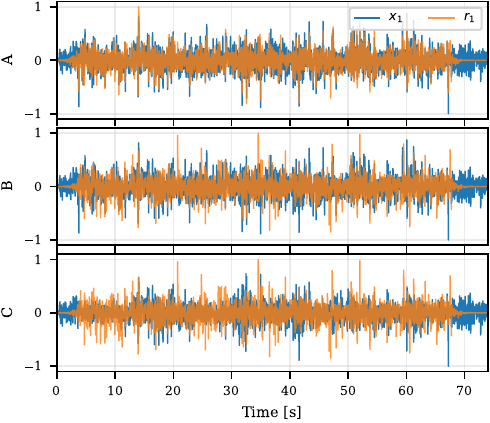}
    \caption{Noisy input and reference signals for Sections A--C, shown for the left channel. Display normalization is applied only to clarify the temporal relationship. The similar envelope behavior indicates that the reference contains information about the train interference observed in the primary audio.}
    \label{fig:input}
\end{figure}

Fig.~\ref{fig:input} displays the left-channel noisy input and the corresponding reference for Sections A--C. The plotted signals are normalized for visualization only, whereas metrics are computed using the original amplitudes. Sections A and B exhibit larger input amplitudes than Section C, indicating stronger interference or higher program energy. The reference envelopes contain temporal structures that coincide with the primary recording, which supports the physical assumption that both signals contain filtered versions of the same real train-noise source. At the same time, the waveforms are not identical, demonstrating the need for adaptive path estimation rather than direct subtraction.

\begin{figure}[!t]
    \centering
    \includegraphics[width=\columnwidth]{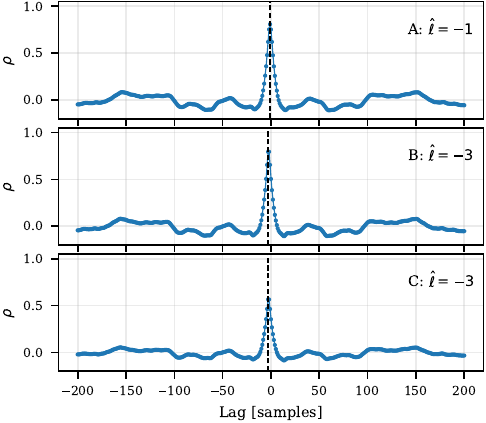}
    \caption{Normalized cross-correlation between the noisy input and the reference. The dashed vertical marker indicates the selected delay used before RLS adaptation. Peaks near zero lag confirm that the reference measurement is strongly related to the real train-noise component.}
    \label{fig:corr}
\end{figure}

The cross-correlation functions in Fig.~\ref{fig:corr} provide direct evidence of reference relevance. Narrow peaks occur close to zero lag in all sections. The selected lags are -1 sample for Section A and -3 samples for Sections B and C. At 11.025 kHz, these corrections correspond to sub-millisecond offsets. Although the shifts are small, alignment is beneficial because the adaptive filter has finite length and should not waste coefficients on deterministic delay compensation.

\subsection{Waveform Behavior After Adaptive Cancellation}
\begin{figure}[!t]
    \centering
    \includegraphics[width=\columnwidth]{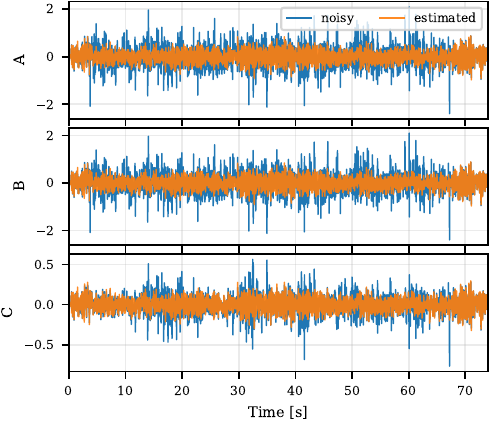}
    \caption{Time-domain comparison between the noisy input and the final estimate for the left channel. The adaptive stage subtracts the component predictable from the reference, and the postfilter smooths high-frequency residual artifacts.}
    \label{fig:time}
\end{figure}

Fig.~\ref{fig:time} compares the noisy waveform with the final RLS-plus-low-pass estimate for the left channel. The processed signals preserve the broad audio envelope while reducing pronounced high-amplitude train-related bursts. Sections A and B show the most visible attenuation because the initial reference correlation and RMS levels are higher. Section C begins with a lower-amplitude disturbance, and its time-domain modification is correspondingly less dramatic. The result is consistent with the expected behavior of an adaptive noise canceller: strong reference-predictable energy is attenuated more visibly than weak or uncorrelated ambience.

The waveform result should be interpreted as an interference-suppression indicator rather than proof of perfect clean-signal recovery. Since the clean target is unavailable, the processed waveform cannot be compared to an exact ground truth. Nevertheless, the reduction of large reference-related fluctuations is consistent with the adaptive cancellation mechanism described by (\ref{eq:mimoerr}). The use of real recordings also makes the waveform comparison meaningful: attenuation occurs under environmental acoustic conditions rather than under an idealized additive-noise simulation.

\subsection{Spectral Evidence}
\begin{figure}[!t]
    \centering
    \includegraphics[width=\columnwidth]{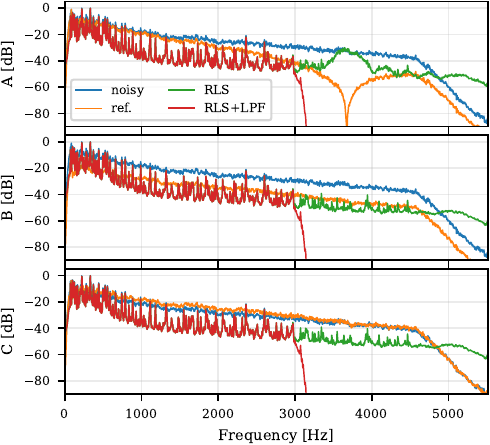}
    \caption{Welch PSD estimates for the left channel. The RLS stage reduces the reference-correlated broadband component, while the low-pass postfilter mainly affects the upper audio band. Spectral normalization is performed independently for each section.}
    \label{fig:spectrum}
\end{figure}

Fig.~\ref{fig:spectrum} shows Welch power spectral density estimates for the noisy input, reference, RLS output, and RLS-plus-low-pass output. The spectra are normalized to the maximum noisy-input PSD in each section to emphasize relative spectral changes. The RLS output exhibits attenuation in frequency regions strongly represented in the reference, indicating removal of train-related components that are coherent across the two measurements. The low-pass postfilter adds additional suppression in the upper band above approximately 3 kHz, matching the design cutoff and reducing residual high-frequency artifacts.

The spectral plots also reveal an important design trade-off. Strong upper-band filtering can reduce residual hiss-like artifacts and high-frequency train components, but desired audio components in the same band may also be attenuated. For future studies, the cutoff frequency should therefore be optimized with perceptual tests, task-dependent quality criteria, or adaptive postfiltering. Classical speech-enhancement methods such as spectral subtraction and minimum mean-square error short-time spectral amplitude estimation provide useful comparison baselines for non-reference postprocessing \cite{boll,ephraim,loizou}.

\subsection{No-Reference Quantitative Indicators}
Table~\ref{tab:metrics} reports RMS changes and residual reference-correlation values. The maximum normalized correlation with the reference decreases from 0.832 to 0.016 in Section A left channel, from 0.667 to 0.014 in Section B left channel, and from 0.487 to 0.014 in Section C left channel. Similar behavior occurs in the right channel. The final residual correlation values are concentrated in the narrow range 0.011--0.016, indicating substantial removal of the component predictable from the reference.

\begin{table}[!t]
\centering
\caption{No-reference performance indicators. RMS values are reported as output/input change in dB. Correlation values are maximum absolute normalized correlations with the corresponding reference channel.}
\label{tab:metrics}
\scriptsize
\begin{tabular}{c c c c c c}
\toprule
Sec. & Lag & RMS change & \multicolumn{2}{c}{\(\eta\) before} & \multicolumn{1}{c}{\(\eta\) after} \\
 & [samp.] & [dB] L/R & L & R & L/R \\
\midrule
A & -1 & -4.80 / -3.25 & 0.832 & 0.722 & 0.016 / 0.016 \\
B & -3 & -4.79 / -3.24 & 0.667 & 0.575 & 0.014 / 0.011 \\
C & -3 & -2.89 / -1.79 & 0.487 & 0.386 & 0.014 / 0.011 \\
\bottomrule
\end{tabular}
\end{table}

\begin{figure}[!t]
    \centering
    \includegraphics[width=0.95\columnwidth]{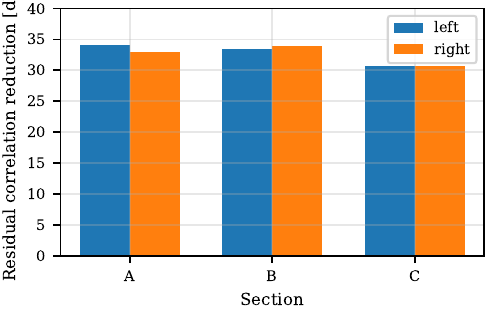}
    \caption{Reduction of the maximum residual normalized correlation with the reference. More negative values indicate stronger attenuation of the real train-noise component predicted from the reference recording.}
    \label{fig:corrred}
\end{figure}

Fig.~\ref{fig:corrred} summarizes the correlation-ratio reduction. The attenuation ranges from about 30.6 dB to 34.1 dB across channels and sections. The stronger absolute RMS decrease in Sections A and B is consistent with the higher initial interference level visible in Fig.~\ref{fig:input}. The lower RMS decrease in Section C should not be interpreted as algorithmic failure; rather, Section C contains a lower-amplitude disturbance and therefore has less removable reference-correlated energy.

\section{Discussion and Importance for Future Studies}
The results demonstrate the practical value of reference-based adaptive filtering for real acoustic interference. The experiment differs from a simple classroom simulation because the disturbing sound is a real train recording embedded in environmental audio. Real recordings contain nonstationarity, spectral overlap, reverberation, and measurement-path mismatch. The effects are exactly the conditions that future audio-denoising and acoustic-monitoring systems must address outside controlled laboratory environments.

The RLS estimator is effective because it exploits the physical relationship between the reference and the disturbance. The method does not require a statistical model of the desired stereo program and does not require supervised training data. The property is important for future studies in scenarios where clean recordings are unavailable, such as outdoor transportation monitoring, mobile recording near railways, hearing-assistive devices in public spaces, and restoration of archival recordings containing transportation noise. A correlated reference microphone or auxiliary environmental channel can provide information that purely single-channel denoising methods cannot access.

The experiment also clarifies several limitations that should guide future developments. First, the reference should be as decorrelated as possible from the desired audio. If program content leaks into the reference path, adaptive cancellation may remove desired components. Second, the linear FIR model may not capture all path variations, especially when source or microphone positions change. Third, anti-causal taps are suitable for offline restoration but not for zero-latency applications. Fourth, the postfilter improves smoothness but introduces a frequency-selective coloration risk. The limitations motivate future comparisons with normalized LMS, affine projection algorithms, subband RLS, frequency-domain adaptive filtering, Kalman filtering, and modern neural denoisers.

Future studies should also include controlled ground-truth experiments. A clean stereo program could be recorded separately and mixed with real train-noise recordings at known levels, enabling SNR, MSE, perceptual evaluation of speech quality, and objective audio-quality metrics. After such controlled validation, the algorithm could be tested again on fully real scenes. Subjective listening should be incorporated through standardized protocols because low residual correlation does not always imply best perceived quality \cite{itur}. A useful long-term direction is a hybrid system in which reference-based RLS removes the physically correlated train component and a perceptual postprocessor removes residual uncorrelated environmental noise.

\section{Reproducibility and Practical Deployment Considerations}
The experimental pipeline is deterministic and can be reproduced from the MATLAB files, the parameter set in Table~\ref{tab:params}, and the reported metric definitions. All objective values are computed before audio-file peak normalization, preventing normalization from artificially changing the quantitative indicators. The same algorithmic parameters are used for the three sections, which avoids overfitting each recording and permits fair comparison between noise conditions. Such reproducibility is important for future studies because adaptive filters can otherwise be tuned informally until a visually pleasing waveform is obtained.

Practical deployment requires several modifications. The anti-causal taps used for offline restoration can be converted into a causal implementation by delaying the primary audio by the number of future reference samples. For the selected configuration, a delay of 15 samples corresponds to approximately 1.36 ms at 11.025 kHz, which is small for many monitoring and postproduction applications. However, full RLS has quadratic complexity in the regressor length because the inverse correlation matrix must be updated at each sample. With 60 reference coefficients per output, offline processing is straightforward, but embedded real-time systems may require fast transversal RLS, frequency-domain adaptive filtering, subband processing, or normalized LMS variants.

Reference-sensor placement is also decisive. A useful reference should capture the real train-noise source with high coherence while minimizing leakage from the desired stereo program. In an outdoor railway environment, a reference microphone closer to the track, an accelerometer on a structural element, or an additional environmental microphone can improve the separability of the target and interference. Multiple references can be beneficial when the train sound reaches the primary microphones through several propagation paths. The two-channel formulation already reflects the idea, because both reference channels contribute to each stereo output estimate.

The experimental protocol provides a practical foundation for future benchmark design. A recommended next step is to combine real train-noise recordings with separately acquired clean stereo material at controlled mixing ratios. Such a semi-controlled dataset would preserve realistic environmental noise while enabling ground-truth metrics. Additional fully real recordings should then be used to test generalization across station platforms, tunnels, open railways, weather conditions, train speeds, and microphone geometries. A hybrid enhancement architecture is also promising: reference-based RLS can remove the coherent train component, while a perceptual postprocessor can address residual uncorrelated ambience.

The reproducibility requirement also has a methodological implication for future experimental studies. Parameter values should be reported together with the signal alignment rule, normalization convention, tap indexing, and metric computation interval. Without these details, two implementations of the same adaptive algorithm can produce different residual waveforms even when the same audio files are used. In the present configuration, the delay estimate is obtained before adaptation, the reference vector contains both past and future samples, and quantitative values are obtained before peak normalization. These conventions make the results auditable and allow later investigations to isolate the influence of one design choice at a time.

Several extensions can be evaluated directly from the same framework. Variable forgetting factors may improve tracking when the train source becomes stronger or weaker during the recording. Diagonal loading and coefficient regularization may increase numerical stability when the stereo reference channels are highly correlated. Subband processing can reduce complexity and permit frequency-selective adaptation, which is especially relevant for train noise because low-frequency rolling components and upper-band friction components may require different adaptation rates. A perceptual weighting stage could additionally prevent the postfilter from attenuating audio components that are important for listening quality. Reporting these variants on the same real recordings would make algorithmic trade-offs visible: convergence speed, residual train-noise suppression, preservation of stereo ambience, and avoidance of musical or muffled artifacts could be assessed under one common protocol.

\section{Conclusion}
Reference-based RLS interference mitigation was evaluated on three real stereo audio recordings containing train noise and environmental background. The signal model treats the primary recording and the reference as different filtered observations of a common external train-noise source. A two-reference/two-output RLS filter estimates the component of the noisy audio that is predictable from the stereo reference; a low-pass FIR postfilter then smooths residual high-frequency artifacts.

The experimental results indicate substantial suppression of reference-related train interference. Across all sections and channels, the maximum normalized correlation with the reference decreases to 0.011--0.016, corresponding to approximately 30.6--34.1 dB correlation-ratio reduction. RMS decreases of 1.8--4.8 dB are observed, with the strongest attenuation in sections containing larger initial disturbances. Waveform and spectral plots support the conclusion that real train-noise components are effectively reduced under environmental recording conditions.

The absence of a clean target prevents claims of true SNR improvement or exact reconstruction accuracy. Nevertheless, the combination of physical signal modeling, residual-correlation reduction, spectral evidence, and real-sound experimentation provides a meaningful foundation for future studies on practical audio interference mitigation. Further progress should include clean-reference validation data, subjective listening tests, comparison with additional adaptive and learning-based baselines, and real-time causal implementations.

\end{document}